\documentclass[12pt,a4paper]{article}

\usepackage[margin=1in]{geometry}
\usepackage{amsmath,amssymb,bm}
\usepackage{graphicx}
\usepackage{hyperref}
\usepackage{setspace}
\usepackage{cite}
\usepackage{setspace}
\begin{document}

\title{\textbf{Structural Relaxation in Supercooled Water:
Hydrogen-Bond Exchange, Defect Propagation,
and Power-Law Scaling}}

\author{
Biman Bagchi\\
Solid State and Structural Chemistry Unit\\
Indian Institute of Science, Bengaluru 560012, India/\\
Email: profbiman@gmail.com
}

\date{}

\maketitle

\begin{abstract}
Molecular simulations of deeply supercooled water
reveal an intriguing power-law relationship between
structural relaxation time and the concentration
of a specific three-coordinated hydrogen-bond defect,
$3'=H^1O^2$. The microscopic origin of this
relationship, and the exceptional dynamical activity
of this defect, remain incompletely understood.
We develop a reaction--transport theory in which
active $3'$ defects propagate through concerted
hydrogen-bond rearrangements. An unsatisfied donor
hydrogen forms a new bond, creating a transient
five-coordinated intermediate; subsequent breaking
of another bond transfers the defect to a neighboring
molecule. This mechanism links molecular jump
reorientation to the migration of coordination defects
through a continuously evolving hydrogen-bond network,
without requiring destruction of its overall
connectivity. The dynamics includes propagation,
self-decay, and encounter-induced annihilation with
five-coordinated defects. Independently acting defects
yield the mean-field result
$\tau_\alpha\propto c_{3'}^{-1}$.
By contrast, concentration-dependent propagation,
defect interactions, and correlations between
successive network rearrangements can generate a
larger, noninteger effective exponent $\alpha_D$.
Analytical results and illustrative calculations
on a four-connected network show how elementary
exchange kinetics, finite defect lifetimes, and
repeated exploration of local environments modify
the relaxation efficiency and hence the observed
power law. Although the elementary mechanism is
specific to water, the underlying reaction--transport
description may extend to other network-forming
liquids with identifiable mobile coordination
defects. The theory thereby connects microscopic
hydrogen-bond rearrangements, coordination-resolved
equilibrium statistics, and the anomalous slowing
down of supercooled water.
\end{abstract}

\maketitle
% ================ INTRODUCTION =================

\section{Introduction}

The molecular dynamics of water presents a fundamental
problem in the statistical mechanics of network-forming
liquids. Water possesses an extended, continuously
reconstructing hydrogen-bond network that strongly
constrains local molecular configurations. Nevertheless,
molecular translation and rotation remain remarkably
rapid under ambient conditions. Upon supercooling,
structural relaxation slows dramatically and exhibits
strongly non-Arrhenius temperature dependence.
Understanding how molecular rearrangements proceed
within an extensively connected hydrogen-bond network,
and why these processes become increasingly sluggish
upon cooling, remains a central challenge.

An important connection between local structural order and the anomalous properties of water was established by Errington and Debenedetti \cite{Errington2001}, who demonstrated the relationship between translational and tetrahedral order and the thermodynamic and transport anomalies of liquid water. Their results emphasize that the unusual dynamics of water cannot be separated from the organization of its hydrogen-bond network. This connection provides a broader motivation for examining how local coordination defects participate in molecular relaxation.

An early and remarkably insightful description
of liquid water as a dynamically connected
network was developed by Julian Gibbs and
coworkers. Gibbs, Cohen, Fleming, and Porosoff
proposed that liquid water could be viewed
as an infinitely branched, randomly connected
gel of hydrogen bonds, with continuous breaking,
formation, and interchange of the network
connections \cite{Gibbs1973}. In this picture,
the melting of ice involves the loss of the
ordered arrangement of hydrogen-bonded rings,
rather than the complete destruction of
hydrogen-bond connectivity. Fleming and Gibbs
subsequently developed statistical-mechanical
lattice models incorporating directional
hydrogen bonding and network connectivity
\cite{FlemingGibbs1974,FlemingGibbs1974II}.
The relationship between gelation theory
and molecular-dynamics descriptions of the
hydrogen-bond network was examined further
by Bagchi, Donoghue, and Gibbs
\cite{BagchiDonoghueGibbs1983}. These early
studies established a useful physical picture
of water as a fluctuating, extensively connected
network whose molecular arrangement changes
continually without necessarily destroying
its overall connectivity.

The anomalous dynamics of supercooled water has been
investigated extensively. Early investigations suggested
changes in the character of relaxation around 228 K
\cite{Ito1999,Bergman2000,Mallamace2012}.
The experimental situation is complicated by
crystallization, which restricts direct measurements
of deeply supercooled bulk water. Nevertheless,
investigations of crystallization kinetics, structural
relaxation, and transport have provided important
information about molecular motion in this region
\cite{Xu2016,Kringle2021,Mussa2023}.
These observations motivate a microscopic description
of water dynamics that explicitly recognizes the
special role of hydrogen-bond rearrangements.

The dynamical significance of this network
picture emerged from the molecular simulations
of Sciortino and coworkers. Sciortino, Poole,
Stanley, and Havlin showed that the lifetimes
of hydrogen bonds and of the connected
network increase strongly upon supercooling,
providing a connection between network
persistence and the anomalous slowing down
of water \cite{Sciortino1990}. In a subsequent
study, Sciortino, Geiger, and Stanley identified
the fifth molecular neighbor and associated
bifurcated hydrogen bonds as defects of the
tetrahedral network, capable of providing
low-energy pathways for its restructuring
\cite{Sciortino1991, Sciortino1990Fifth}. Their influential
Nature paper demonstrated that these defects
can facilitate molecular mobility by acting
as local catalysts for rearrangements between
tetrahedral configurations
\cite{Sciortino1991}. A more detailed analysis
further established the connection between
topological network defects and molecular
mobility \cite{Sciortino1992}. These studies
introduced an important microscopic insight:
molecular relaxation in water can be promoted
by the motion and rearrangement of local
coordination defects, without requiring
the hydrogen-bond network as a whole to
disintegrate.

A major advance in understanding molecular
reorientation was provided by Laage and Hynes
\cite{Laage2006}, who established that water
molecules undergo large-amplitude angular jumps
through hydrogen-bond partner exchange. This
mechanism revealed the importance of concerted
bond breaking and formation in molecular rotation.
Subsequently, Jana, Singh, and Bagchi
\cite{Jana2011} demonstrated string-like propagation
of five-coordinated defect states in supercooled
water. These studies suggest a close connection
between local hydrogen-bond exchange, the propagation
of coordination excitations, and correlated
molecular rearrangements.

A complementary perspective on the molecular
dynamics of supercooled water has emerged from
the studies of Daschakraborty and coworkers.
Using molecular dynamics simulations and a
translational jump-diffusion description,
Dueby, Dubey, and Daschakraborty demonstrated
the increasing importance of intermittent
translational jumps in the breakdown of the
Stokes--Einstein relation upon supercooling
\cite{Dueby2019}. A subsequent perspective
discussed the broader implications of this
mechanism for the dynamical anomalies of water
\cite{Dubey2021}. More recently, Nagaraj and
Daschakraborty identified a structural sequence
preceding translational jumps, involving a
loss of local tetrahedral order, weakening
of hydrogen bonds, and correlated changes
in the first and second solvation shells
\cite{Nagaraj2026}. These findings emphasize
the role of localized but cooperative
hydrogen-bond network rearrangements in
molecular transport. Although translational
jumps of individual molecules and the
propagation of coordination defects are
distinct dynamical processes, both may
originate from related local network
reorganizations. Establishing their
microscopic connection remains an
interesting question.

A particularly intriguing development emerged from
the molecular simulations of Saito, Bagchi, and
Ohmine \cite{Saito2018}. They found that extended
defect-rich regions progressively fragment into
smaller clusters and isolated defects upon deep
supercooling. Nevertheless, structural relaxation
persists, with an important contribution from a
particular three-coordinated species, denoted here
by $3'=H^1O^2$, possessing one donated and two
accepted hydrogen bonds.

Even more striking is the reported power-law
relationship between structural relaxation time
and the concentration of this active defect.
We write this relationship as
\begin{equation}
\tau_\alpha(T)\sim
[c_{3'}(T)]^{-\alpha_D}.
\label{eq:intro_power_law}
\end{equation}
The simulations suggest an unusually large
effective exponent, approximately $4.3$, over
the investigated temperature range
\cite{Saito2018}. This exponent need not be
universal or asymptotic. Nevertheless, the
observation raises an important theoretical
question: why should the relaxation time
depend so strongly on the population of a
specific coordination defect?

The broader physical significance of these observations
has been emphasized by Ohmine and Saito
\cite{OhmineSaito2021}. Their analysis identifies
the unusual local geometry of the $H^1O^2$
defect, its enhanced vibrational activity in the
$300$--$400\,{\rm cm}^{-1}$ region, and its
coupling to collective hydrogen-bond network
rearrangements. They also point out the
participation of the same coordination species
in ice melting. These findings suggest that
the remarkable activity of the $3'$ defect
originates not simply from its incomplete
coordination but from the ease with which
its local environment can undergo concerted
bond exchanges.

The distinction between the geometrical
connectivity of defective regions and the
dynamical propagation of individual defects
is crucial. A defect-rich cluster consists
of simultaneously neighboring defective
environments. By contrast, a coordination
excitation can propagate through successive
hydrogen-bond rearrangements involving
molecules that were initially four-coordinated.
Fragmentation of the defective subnetwork
therefore need not prevent defect-mediated
relaxation through the underlying
hydrogen-bond network.

In the present work, we propose a microscopic
mechanism for the exceptional dynamical
activity of the $3'$ defect. Its unsatisfied
OH donor can form an additional hydrogen
bond with a suitably oriented neighboring
molecule, producing a transient
five-coordinated intermediate. Subsequent
breaking of another hydrogen bond restores
fourfold coordination at the intermediate
site while transferring the three-coordinated
defect to a different molecule. The resulting
process is a concerted rearrangement involving
three water molecules. It provides a possible
microscopic connection between the molecular
jump mechanism of Laage and Hynes, the
propagation phenomena studied by Jana
and coworkers, and the persistent relaxation
associated with active three-coordinated
defects in the simulations of Saito,
Bagchi, and Ohmine.

The donor--acceptor asymmetry of the
three-coordinated species is important
in this mechanism. Unlike the alternative
three-coordinated configuration with two
donated and one accepted hydrogen bond,
the $3'$ species possesses an unsatisfied
donor that can initiate this particular
exchange pathway. Intermolecular
vibrational motions, involving coupled
hydrogen-bond stretching and hindered
molecular rotations, may facilitate
the formation of the transient
five-coordinated intermediate.

The propagating excitation has a finite
lifetime. It can disappear through local
network reconstruction or through
annihilation following an encounter
with an existing five-coordinated defect.
The latter must be distinguished from
the transient five-coordinated intermediate
formed during a successful propagation
event. The competition between propagation
and disappearance determines the number
and spatial extent of rearrangements
generated by an individual excitation.

These considerations provide a physical
basis for understanding the unusual
power-law relaxation. In a homogeneous
mean-field description, independently
acting mobile defects produce a
relaxation time inversely proportional
to their concentration, corresponding
to an exponent of unity. However,
hydrogen-bond exchange kinetics,
interactions between defects, and
correlations among their propagation
trajectories modify the productive
relaxation activity of each excitation.
The resulting exponent can therefore
exceed unity and vary with temperature.
Our objective is to establish this
connection analytically and examine
its consequences in a minimal
network model, without assuming a
particular exponent in advance.

A quantitative theory must also address
the equilibrium population of active
coordination defects. In a recently
developed statistical-mechanical theory
of supercooled water, Bagchi introduced
a collective description of defective
hydrogen-bond environments, combining
their configurational entropy with
cooperative stabilization of the
tetrahedral network
\cite{BagchiDefect2026}.
The resulting thermodynamics determines
the equilibrium concentration of
collective defective environments
and its coupling to density.

This collective defect concentration
is not identical to the population
of active $3'$ excitations. The former
includes different coordination states
and nonideal molecular environments,
whereas the latter identifies a
specific donor--acceptor configuration.
Connecting the two descriptions
requires coordination-resolved
statistical mechanics consistent
with the underlying defect entropy.
This connection is important for
relating equilibrium network structure
to the microscopic processes
responsible for relaxation.

The broader theoretical perspective
is provided by collective descriptions
of liquid dynamics. Conventional
mode-coupling approaches have been
applied to network-forming liquids,
including silica \cite{SciortinoKob2001}.
Nonlinear fluctuating hydrodynamics
incorporates the conserved density
and momentum fields
\cite{DasMazenko1986}, while
phenomenological defect variables
have also been considered in
related formulations
\cite{YeoMazenko1995}.
The present approach differs in
starting from chemically identifiable
coordination defects and their
microscopic propagation and reaction
mechanisms. Establishing these
processes provides a foundation
for a subsequent mode-coupling-like
theory in which defect dynamics
and collective hydrodynamic
fluctuations are treated consistently.

The paper is organized as follows.
Section II introduces the
hydrogen-bond network and its
coordination states. Section III
develops the propagation and
reaction dynamics of mobile
defects. Section IV establishes
the mean-field power law and
the role of network correlations.
Section V examines the microscopic
three-molecule exchange mechanism
and its connection to defect
statistical mechanics. Section VI
presents analytical and numerical
results for the effective
relaxation exponent. We conclude
with the implications for
collective relaxation in
network-forming liquids and
the development of a more
complete dynamical theory.

%=================  SECTION 2 =====================
% =======================  Section =========================

\section{Hydrogen-Bond Network, Defect States,
and Elementary Rearrangements}
\label{sec:network}

We begin by constructing a microscopic description
of the hydrogen-bond network and its coordination
defects. The distinction between donated and
accepted hydrogen bonds is essential for identifying
the active defects and their possible rearrangements.
We first define the network and its equilibrium
defect populations, and then establish the elementary
bond-exchange processes and statistical constraints
that govern their dynamics.

\subsection{The evolving hydrogen-bond network}

We regard liquid water as a fluctuating
hydrogen-bond network whose vertices are
water molecules and whose edges represent
hydrogen bonds. The network is continuously
reconstructed through bond formation, breaking,
and exchange. Unlike a permanent network,
its connectivity is therefore a dynamical
variable. We consider a system containing
$N$ water molecules.

Hydrogen bonds are directional, and it is
essential to distinguish the donor and
acceptor molecules. We introduce the
directed hydrogen-bond indicator in terms of the usual theta functions,

\begin{equation}
h_{ij}(t)=
\Theta\left(r_c-r_{ij}(t)\right)
\Theta\left(\theta_c-\theta_{ij}^{D}(t)\right),
\label{eq:network_hij}
\end{equation}

Here $h_{ij}(t)=1$ when molecule $i$ donates
a hydrogen bond to molecule $j$, and zero
otherwise. The quantity $r_{ij}(t)$ is the
instantaneous oxygen--oxygen separation,
while $r_c$ is the corresponding cutoff
distance. The angle $\theta_{ij}^{D}(t)$
characterizes the orientation of the donor
hydrogen relative to the acceptor oxygen,
and $\theta_c$ is the angular cutoff.
The function $\Theta(x)$ is the Heaviside
step function, equal to unity for positive
$x$ and zero for negative $x$.

Since each water molecule possesses two
donor hydrogens, the angular criterion
must be applied using a definite
hydrogen-assignment convention. The same
convention is used throughout the
calculation of equilibrium coordination
populations and dynamical correlations.

When more than one donor hydrogen is
geometrically eligible, an unambiguous
bond-assignment convention must be adopted.
The same convention must be used in
evaluating the equilibrium populations
and their dynamical correlations.

The corresponding undirected adjacency
matrix is
\begin{equation}
b_{ij}(t)=
1-\left[1-h_{ij}(t)\right]
  \left[1-h_{ji}(t)\right].
\label{eq:network_bij}
\end{equation}
Thus $b_{ij}=b_{ji}$.

Thus, $b_{ij}=b_{ji}$ specifies the
undirected connectivity of the network,
whereas $h_{ij}$ retains the direction
of hydrogen-bond donation. We adopt a
bond-assignment convention that excludes
simultaneous donation in both directions
between the same molecular pair, so that
$h_{ij}h_{ji}=0$.

We restrict the elementary graph
description to configurations with
at most one directed hydrogen bond
between a given molecular pair.
Consequently, $h_{ij}h_{ji}=0$.

The numbers of donated and accepted
hydrogen bonds of molecule $i$ are
\begin{equation}
n_i^D(t)=\sum_j h_{ij}(t),
\qquad
n_i^A(t)=\sum_j h_{ji}(t).
\label{eq:network_DA}
\end{equation}

The total coordination number is then
\begin{equation}
z_i(t)=n_i^D(t)+n_i^A(t)
      =\sum_j b_{ij}(t).
\label{eq:network_zi}
\end{equation}

The donor--acceptor resolution is
essential because two molecules
having the same coordination number
may exhibit different dynamical
properties.

Each hydrogen bond contributes
one donated and one accepted
bond to the network. Therefore,
\begin{equation}
\sum_i n_i^D(t)
=
\sum_i n_i^A(t)
=
N_{\rm HB}(t),
\label{eq:network_balance}
\end{equation}
where $N_{\rm HB}(t)$ is the
instantaneous number of hydrogen
bonds.

Equivalently,
\begin{equation}
\sum_i
\left[n_i^D(t)-n_i^A(t)\right]=0.
\label{eq:network_balance_difference}
\end{equation}

Equations~(\ref{eq:network_balance})
and (\ref{eq:network_balance_difference})
provide important constraints on
the allowed elementary reactions.

The total number of hydrogen bonds
need not be conserved during
bond breaking and formation.
However, the equality of the
total numbers of donated and
accepted bonds must hold for
every network configuration.

\subsection{Collective defective environments
and coordination-resolved defects}

We retain the definition of the
collective defective environment
introduced in our thermodynamic
description of supercooled water
\cite{BagchiDefect2026}.

A molecule belongs to an ideal
environment when it is
four-coordinated and all its
hydrogen-bonded neighbors are
also four-coordinated.

The corresponding indicator is
\begin{equation}
I_i(t)=
\delta_{z_i(t),4}
\prod_j
\left[
1-b_{ij}(t)
+b_{ij}(t)\delta_{z_j(t),4}
\right].
\label{eq:network_Ii}
\end{equation}

The defective-environment
indicator is
\begin{equation}
d_i(t)=1-I_i(t).
\label{eq:network_di}
\end{equation}

The collective defect concentration is
\begin{equation}
c_d(t)=
\frac{1}{N}\sum_i d_i(t),
\label{eq:network_cd}
\end{equation}
with complementary ideal fraction
\begin{equation}
c_I(t)=1-c_d(t).
\label{eq:network_cI}
\end{equation}

This definition is deliberately
the same as that employed in the
defect thermodynamics. In
particular, a four-coordinated
molecule may be classified as
defective when its neighboring
environment fails the
four-coordination criterion.

The definition is based on
coordination and local
connectivity. It does not,
by itself, require that every
four-coordinated molecule have
the donor--acceptor state
$(n_i^D,n_i^A)=(2,2)$.
The latter is an additional
chemical distinction retained
in the coordination-resolved
description.

For the dynamics, we must
distinguish the specific
coordination species from
the broader variable $c_d$.

The active three-coordinated
species is denoted by $3'$,
corresponding to the
donor--acceptor state
\begin{equation}
3'\equiv H^1O^2
\equiv (n^D,n^A)=(1,2).
\label{eq:network_3prime}
\end{equation}

Its microscopic indicator is
\begin{equation}
a_i(t)=
\delta_{n_i^D(t),1}
\delta_{n_i^A(t),2},
\label{eq:network_ai}
\end{equation}
and its instantaneous
concentration is
\begin{equation}
c_{3'}(t)=
\frac{1}{N}\sum_i a_i(t).
\label{eq:network_c3}
\end{equation}

The other common
three-coordinated
configuration is
\begin{equation}
H^2O^1
\equiv(n^D,n^A)=(2,1).
\label{eq:network_other3}
\end{equation}

The distinction between
these species is important
because their dynamics
need not be equivalent.

For five-coordinated
molecules we focus on
the configuration
\begin{equation}
5\equiv H^2O^3
\equiv(n^D,n^A)=(2,3),
\label{eq:network_5}
\end{equation}
with indicator
\begin{equation}
f_i(t)=
\delta_{n_i^D(t),2}
\delta_{n_i^A(t),3},
\label{eq:network_fi}
\end{equation}
and concentration
\begin{equation}
c_5(t)=
\frac{1}{N}\sum_i f_i(t).
\label{eq:network_c5}
\end{equation}

A molecule in either
the $3'$ or the
five-coordinated state
necessarily belongs
to the collective
defective population.

At every instant,
\begin{equation}
a_i\leq d_i,
\qquad
f_i\leq d_i,
\label{eq:network_inclusion}
\end{equation}
and hence
\begin{equation}
c_{3'}+c_5\leq c_d.
\label{eq:network_population_bound}
\end{equation}

The inequality follows
because the two
coordination states are
mutually exclusive, while
$c_d$ also contains other
defective environments.

Thus $c_d$, $c_{3'}$,
and $c_5$ are related
structural observables,
not independently
defined thermodynamic
populations.

\subsection{Equilibrium thermodynamics
and network-state statistics}

The equilibrium concentration
of collective defective
environments is constrained
by the defect free energy
developed previously
\cite{BagchiDefect2026}.

For reference, the reduced
uniform defect contribution
per molecule is
\begin{align}
f_d(c_d)={}&
k_BT\left[
c_d\ln c_d+
(1-c_d)\ln(1-c_d)
-c_d\ln g_d
\right]
\nonumber\\
&+\epsilon_d c_d
-J(1-c_d)^2.
\label{eq:network_fd}
\end{align}

The equilibrium condition is
\begin{equation}
k_BT\ln
\frac{c_d^{\rm eq}}
{g_d(1-c_d^{\rm eq})}
+\epsilon_d
+2J(1-c_d^{\rm eq})=0.
\label{eq:network_equilibrium_cd}
\end{equation}

This expression represents
the reduced defect sector.
In the complete theory,
the equilibrium defect
concentration and
specific volume are
determined together
from the coupled Gibbs
free energy.

No additional relation
between density and
defect concentration
is introduced here.

The thermodynamic free
energy determines the
equilibrium concentration
of defective environments,
but does not uniquely
specify the statistical
weights of individual
donor--acceptor
coordination species.

Let ${\cal C}$ denote
a complete configuration
of the directed
hydrogen-bond network.

We introduce its
normalized equilibrium
probability
\begin{equation}
\Pi_{\rm eq}({\cal C};T,P),
\qquad
\sum_{\cal C}
\Pi_{\rm eq}({\cal C};T,P)=1.
\label{eq:network_Pi}
\end{equation}

A coarse-grained
network configuration
may possess internal
positional and
orientational
multiplicity. Its
equilibrium probability
must include those
statistical weights.

The average collective
defect concentration
is obtained from
\begin{equation}
c_d^{\rm eq}
=
\frac{1}{N}
\sum_{\cal C}
\Pi_{\rm eq}({\cal C})
\sum_i d_i({\cal C}).
\label{eq:network_cd_ensemble}
\end{equation}

Similarly,
\begin{equation}
c_{3'}^{\rm eq}
=
\frac{1}{N}
\sum_{\cal C}
\Pi_{\rm eq}({\cal C})
\sum_i a_i({\cal C}),
\label{eq:network_c3_ensemble}
\end{equation}
and
\begin{equation}
c_5^{\rm eq}
=
\frac{1}{N}
\sum_{\cal C}
\Pi_{\rm eq}({\cal C})
\sum_i f_i({\cal C}).
\label{eq:network_c5_ensemble}
\end{equation}

The same network ensemble
must reproduce the
collective defect
thermodynamics in the
appropriate coarse-grained
limit.

This requirement is
important for avoiding
double counting.

The configurational
multiplicity represented
by $g_d$ already
contributes to the
thermodynamic free energy.
It must not be counted
again as an independent
entropy of the resolved
coordination species
or defect clusters.

Instead, a more
microscopic network
model should resolve
the structural
configurations already
represented collectively
by $f_d(c_d)$.

We emphasize that
Eq.~(\ref{eq:network_fd})
alone does not determine
the full probability
$\Pi_{\rm eq}({\cal C})$.

Additional structural
information, particularly
the correlations of
donor--acceptor states,
is required.

\subsection{Elementary bond events
and propagation of a coordination defect}

The elementary dynamics
consists of hydrogen-bond
breaking, formation,
and exchange.

In terms of the directed
bond variables,
a breaking event is
\begin{equation}
h_{ij}:1\longrightarrow0,
\label{eq:network_break}
\end{equation}
and a formation event is
\begin{equation}
h_{ij}:0\longrightarrow1.
\label{eq:network_form}
\end{equation}

Bond breaking decreases
the donated coordination
of molecule $i$ and
the accepted coordination
of molecule $j$ by one.

Bond formation produces
the reverse changes.

A bond-exchange event
may conserve the total
number of bonds while
changing their molecular
connectivity.

An important example is
the transfer of a donated
bond between two donor
molecules.

Suppose molecule $j$
initially donates a
hydrogen bond to
molecule $k$, and
molecule $i$ can
replace it as the
donor to $k$.

The net bond exchange is
\begin{equation}
h_{jk}:1\longrightarrow0,
\qquad
h_{ik}:0\longrightarrow1.
\label{eq:network_donor_exchange}
\end{equation}

This transition conserves
the total number of
hydrogen bonds. It also
leaves the accepted
coordination of
molecule $k$ unchanged.

The donated coordination
of molecule $j$ decreases
by one, while that of
molecule $i$ increases
by one.

For the initial states
\begin{equation}
i:(1,2),
\qquad
j:(2,2),
\label{eq:network_initial_transfer}
\end{equation}
the exchange gives
\begin{equation}
i:(2,2),
\qquad
j:(1,2).
\label{eq:network_final_transfer}
\end{equation}

Thus,
\begin{equation}
3'_i+4_j
\longrightarrow
4_i+3'_j.
\label{eq:network_3prime_transfer}
\end{equation}

This supplies one
microscopically
consistent realization
of the propagation
of a $3'$ coordination
defect.

The notation $4$
in this example
specifically denotes
the $(2,2)$
coordination state.

The transfer requires
the geometry and
orientational
configuration to
permit the new
hydrogen bond.
It is therefore
not assumed to occur
between arbitrary
four-coordinated
molecules.

Furthermore, the
intermediate stages of
the exchange may involve
transient bond breaking
or bifurcated
configurations.

Equation~(\ref{eq:network_donor_exchange})
describes the net
rearrangement rather
than asserting that
the two changes occur
simultaneously.

The defect has moved
from molecule $i$
to molecule $j$,
although neither
molecule needs to
undergo a translation
equal to the
defect-propagation
distance.

The acceptor $k$
participates in
the elementary
rearrangement.
Consequently, the
effective propagation
path need not coincide
with an edge directly
connecting $i$ and $j$
in the instantaneous
hydrogen-bond graph.

This distinction will
be important in
constructing the
propagation operator
in the next section.

\subsection{Microscopic constraints
on defect disappearance}

The same bookkeeping
is necessary when
describing the
disappearance of
active coordination
defects.

A conversion of a
$3'$ state into an
inactive state is
not necessarily
the elimination
of a three-coordinated
molecule.

It may instead
represent conversion
to a different
coordination species
or a local network
reconstruction that
removes the active
excitation.

Similarly, an encounter
between a $3'$ and a
five-coordinated
molecule may remove
both coordination
anomalies, but the
reaction must respect
donor--acceptor balance.

For example, the two
initial configurations
\begin{equation}
H^1O^2+H^2O^3
\label{eq:network_initial_ann}
\end{equation}
contain three donated
and five accepted bonds
when the two molecular
coordination counts
are added.

They cannot be converted
into two $H^2O^2$
states by changing
only bonds within
this pair, because
the latter states
contain four donated
and four accepted bonds.

One possible
coordination-balanced
net outcome is
\begin{equation}
H^1O^2+H^2O^3
\longrightarrow
H^1O^3+H^2O^2.
\label{eq:network_balanced_ann}
\end{equation}

This may be realized
by transferring
one accepted bond
from the five-coordinated
molecule to the
three-coordinated
molecule, with the
external donor of
that bond unchanged.

The resulting molecules
are both four-coordinated,
although one has the
nonstandard
donor--acceptor
state $(1,3)$.

Equation~(\ref{eq:network_balanced_ann})
is an example of a
coordination-consistent
network transition,
not an assertion that
it is the dominant
annihilation pathway
in water.

Formation of two
ordinary $(2,2)$
environments from
the original pair
requires compensating
donor--acceptor changes
at additional network
sites.

Therefore, the
schematic disappearance
of a $3'$ excitation
must ultimately be
expressed in terms
of permitted changes
in the directed
hydrogen-bond graph.

The classification
of such events into
self-decay and
encounter annihilation
will be developed
in the next section.

\subsection{Network transition operator
and detailed balance}

Let ${\cal C}$ specify
a network configuration
and let
\begin{equation}
W({\cal C}\rightarrow{\cal C}';T)
\label{eq:network_W}
\end{equation}
be the rate of an
allowed elementary
transition.

For an effective
Markovian description,
the configuration
probability satisfies
\begin{align}
\frac{\partial P({\cal C},t)}{\partial t}
={}&
\sum_{{\cal C}'}
\big[
W({\cal C}'\rightarrow{\cal C})
P({\cal C}',t)
\nonumber\\
&-
W({\cal C}\rightarrow{\cal C}')
P({\cal C},t)
\big].
\label{eq:network_master}
\end{align}

The transition rates
must respect the
equilibrium distribution
and microscopic
reversibility.

For a reversible
coarse-grained
network dynamics,
we impose
\begin{align}
&\Pi_{\rm eq}({\cal C})
W({\cal C}\rightarrow{\cal C}')
\nonumber\\
&\qquad=
\Pi_{\rm eq}({\cal C}')
W({\cal C}'\rightarrow{\cal C}).
\label{eq:network_DB}
\end{align}

This condition fixes
the ratio of forward
and reverse rates
once the equilibrium
statistical weights
are specified.

It does not determine
their absolute
magnitudes.

A convenient
activated-rate
representation is
\begin{align}
W({\cal C}\rightarrow{\cal C}')
={}&
\nu_{{\cal C}{\cal C}'}
\exp\left\{
-\beta\left[
{\cal G}^{\ddagger}_{{\cal C}{\cal C}'}
-{\cal G}({\cal C})
\right]
\right\},
\label{eq:network_W_activated}
\end{align}
where
\begin{equation}
\Pi_{\rm eq}({\cal C})
\propto
\exp[-\beta{\cal G}({\cal C})].
\label{eq:network_Gweight}
\end{equation}

A symmetric transition-state
free energy and compatible
forward--reverse prefactors
ensure detailed balance.

This representation
is an effective
coarse-grained
rate theory.
The relevant
activation free
energies are
not determined
by the collective
defect free energy
alone.

For an elementary
network event,
the change in the
number of collective
defective environments
is
\begin{equation}
\Delta N_d({\cal C},{\cal C}')
=
N_d({\cal C}')-N_d({\cal C}),
\label{eq:network_DNd}
\end{equation}
with
\begin{equation}
N_d({\cal C})=
\sum_i d_i({\cal C}).
\label{eq:network_Nd}
\end{equation}

The corresponding
collective evolution is
\begin{align}
\frac{d\langle N_d\rangle}{dt}
={}&
\sum_{{\cal C},{\cal C}'}
P({\cal C},t)
W({\cal C}\rightarrow{\cal C}')
\nonumber\\
&\times
\Delta N_d({\cal C},{\cal C}').
\label{eq:network_dNd}
\end{align}

A single hydrogen-bond
event may change the
classification of
several neighboring
environments.

Thus its contribution
to $N_d$ need not be
restricted to
$\Delta N_d=\pm1$.

The total creation,
destruction, and
propagation of
coordination defects
must all be derived
from the same
network transition
operator.

They should not be
introduced as
independent stochastic
processes when they
describe the same
underlying hydrogen-bond
rearrangement.

\subsection{Connectivity and the
dynamical propagation network}

The instantaneous
hydrogen-bond network
is described by
$b_{ij}(t)$.

The subnetwork of
collective defective
environments is
\begin{equation}
B_{ij}(t)
=
d_i(t)b_{ij}(t)d_j(t).
\label{eq:network_Bij}
\end{equation}

Its connected
components characterize
the geometrical
clustering of
defective environments.

The fragmentation
of these clusters
upon cooling,
observed by Saito,
Bagchi, and Ohmine
\cite{Saito2018},
does not imply
fragmentation of
the complete
hydrogen-bond network.

Nor does it
necessarily prevent
the motion of
an isolated active
coordination defect.

The propagation
of $3'$ is governed
by the availability
of allowed
bond-exchange
pathways.

We therefore
distinguish the
instantaneous
hydrogen-bond graph
from the graph of
allowed excitation
transfers.

For a given
network configuration
${\cal C}$, let
\begin{equation}
w_{ij}({\cal C};T)
\label{eq:network_hop_rate}
\end{equation}
denote the rate for
transfer of an active
$3'$ excitation from
site $i$ to site $j$,
conditional on the
local network state.

This rate vanishes
when no permitted
bond rearrangement
can realize the
transfer.

The set of nonzero
$w_{ij}$ defines
the instantaneous
propagation network
for the active
excitation.

This graph need not
be identical either
to $b_{ij}$ or to
the defective
subgraph $B_{ij}$.

The propagation
network is also
dynamic: a successful
exchange can alter
the connectivity
and the subsequent
available transitions.

Thus we must
distinguish three
related but different
structures: the
complete hydrogen-bond
network, the connected
clusters of defective
environments, and
the allowed
propagation pathways
of active coordination
defects.

The corresponding
time-dependent
reaction and
propagation dynamics
will be considered
in the next section.

% ===========================  Section 3 ====================

\section{Propagation and Annihilation of
Mobile Coordination Defects}
\label{sec:propagation}

Having defined the elementary hydrogen-bond
rearrangements, we now examine how an active
three-coordinated defect propagates through the
network. An important feature is that the defect
can move without corresponding translational
motion of an individual water molecule. We develop
a reaction--transport description incorporating
propagation, self-decay, and encounter annihilation,
and obtain the survival probability, lifetime,
and propagation activity of the excitation.

\subsection{Defect propagation and termination}

We consider the motion of an active $3'$ defect
through the hydrogen-bond network. As discussed
in Sec.~\ref{sec:network}, propagation occurs
through successive donor--acceptor bond exchanges,
schematically represented by
\begin{equation}
3'_i+4_j\longrightarrow4_i+3'_j.
\label{eq:three_propagation}
\end{equation}
Each successful transfer can change molecular
orientations and hydrogen-bond partners. The
motion of the coordination defect is therefore
distinct from the translational diffusion of
an individual water molecule.

We distinguish two mechanisms for termination
of the propagating excitation. The first is
self-decay through a local network-reconstruction
event that removes the active $3'$ state without
requiring a five-coordinated reaction partner.
The second is annihilation following an encounter
with a suitable five-coordinated defect.

Both mechanisms are understood as effective
outcomes of allowed microscopic bond
rearrangements, subject to the donor--acceptor
constraints established in Sec.~\ref{sec:network}.
Their rates depend on temperature and local
network configuration. We assume that these
two channels exhaust the termination events
of the excitation being followed.

\subsection{Reactive propagation on the network}

Let $p_i(t)$ be the probability that an initially
identified $3'$ excitation occupies molecular
site $i$ at time $t$ without having undergone
either termination process. Thus
$\sum_i p_i(t)$ is generally less than unity.

For a prescribed effective network, let
$w_{ij}(T)$ denote the rate of successful
transfer from site $i$ to site $j$. These
rates are nonzero only for allowed
bond-exchange pathways. They need not
follow the edges of the instantaneous
undirected hydrogen-bond graph.

We denote the local self-decay and
encounter-annihilation hazards by
$k_s(i,T)$ and $k_a(i,T)$, respectively.
The dilute-excitation master equation is
\begin{align}
\frac{dp_i(t)}{dt}
={}&\sum_{j\ne i}
\left[w_{ji}p_j(t)-w_{ij}p_i(t)\right]
\nonumber\\
&-\left[k_s(i,T)+k_a(i,T)\right]p_i(t).
\label{eq:three_master}
\end{align}

This is a reduced description obtained
from the network transition dynamics
of Sec.~\ref{sec:network}. It is not
an independent set of chemical reactions
added to the microscopic network generator.

Writing $\mathbf p(t)=\{p_i(t)\}$,
we introduce the transient generator
$\mathbf Q$, with
\begin{equation}
Q_{ij}=w_{ji}\quad (i\ne j),
\label{eq:three_Qoff}
\end{equation}
and
\begin{equation}
Q_{ii}=-\sum_{j\ne i}w_{ij}
       -k_s(i,T)-k_a(i,T).
\label{eq:three_Qdiag}
\end{equation}
Hence
\begin{equation}
\frac{d\mathbf p}{dt}=\mathbf Q\mathbf p,
\qquad
\mathbf p(t)=e^{\mathbf Qt}\mathbf p(0).
\label{eq:three_propagator}
\end{equation}

Here the effective transition rates are held
constant during the calculation. This does
not mean that hydrogen bonds remain fixed:
their exchange is already represented by
the allowed excitation transfers.

\subsection{Survival and competing annihilation
mechanisms}

The survival probability of the initially
identified excitation is
\begin{equation}
S_{3'}(t)=
\mathbf 1^{\mathsf T}
e^{\mathbf Qt}\mathbf p(0),
\label{eq:three_survival}
\end{equation}
where $\sum_i p_i(0)=1$.

The probability densities for termination
through the two channels are
\begin{align}
\psi_s(t)&=\sum_i k_s(i,T)p_i(t),
\label{eq:three_psi_s}\\
\psi_a(t)&=\sum_i k_a(i,T)p_i(t).
\label{eq:three_psi_a}
\end{align}

Conservation of probability gives
\begin{equation}
-\frac{dS_{3'}(t)}{dt}
=\psi_s(t)+\psi_a(t).
\label{eq:three_survival_balance}
\end{equation}

The corresponding termination probabilities
are
\begin{equation}
P_s=\int_0^\infty\psi_s(t)\,dt,
\qquad
P_a=\int_0^\infty\psi_a(t)\,dt.
\label{eq:three_branching}
\end{equation}
If every excitation eventually terminates,
\begin{equation}
P_s+P_a=1.
\label{eq:three_branching_sum}
\end{equation}

The mean lifetime is
\begin{equation}
\langle\tau_{3'}\rangle
=\int_0^\infty S_{3'}(t)\,dt
=-\mathbf 1^{\mathsf T}
\mathbf Q^{-1}\mathbf p(0),
\label{eq:three_lifetime}
\end{equation}
provided the transient generator is
invertible and absorption is certain.

These expressions determine not only
the lifetime but also the relative
importance of the two annihilation
mechanisms as functions of temperature.

\subsection{Propagation activity and mean-field limit}

The number of propagation events before
termination measures the dynamical
activity of an individual excitation.
Its mean value is
\begin{equation}
\langle n_{\rm hop}\rangle
=\int_0^\infty dt\,
\sum_i p_i(t)\sum_{j\ne i}w_{ij}.
\label{eq:three_hops}
\end{equation}

This quantity differs from the number
of distinct molecular sites visited,
because the excitation may revisit
previously explored sites. The latter
quantity, together with the spatial
extent of its trajectory, will be
calculated numerically.

For a homogeneous mean-field approximation,
we write
\begin{equation}
k_s(i,T)\simeq k_s(T),
\qquad
k_a(i,T)\simeq k_{\rm enc}(T)c_5(T),
\label{eq:three_mean_field_rates}
\end{equation}
where $k_{\rm enc}$ is an effective
encounter coefficient incorporating
the mean reaction probability and
accessibility of five-coordinated defects.

The resulting survival probability is
\begin{equation}
S_{3'}^{\rm MF}(t)
=\exp[-\Gamma(T)t],
\label{eq:three_MF_survival}
\end{equation}
with
\begin{equation}
\Gamma(T)=k_s(T)+k_{\rm enc}(T)c_5(T).
\label{eq:three_Gamma}
\end{equation}

Consequently,
\begin{equation}
\tau_{3'}^{\rm MF}(T)=\frac{1}{\Gamma(T)},
\label{eq:three_MF_lifetime}
\end{equation}
and
\begin{equation}
P_s^{\rm MF}=\frac{k_s}{\Gamma},
\qquad
P_a^{\rm MF}=\frac{k_{\rm enc}c_5}{\Gamma}.
\label{eq:three_MF_branching}
\end{equation}

If $k_h(T)$ is the uniform total rate
of successful propagation steps,
\begin{equation}
\langle n_{\rm hop}\rangle_{\rm MF}
=\frac{k_h(T)}{\Gamma(T)}.
\label{eq:three_MF_hops}
\end{equation}

Thus the catalytic activity of a
single excitation is governed by
the competition between propagation
and annihilation.

The homogeneous approximation neglects
spatial correlations and repeated
encounters. In particular, the
five-coordinated population alone
does not determine the encounter rate.
When annihilation is calculated explicitly
from network trajectories, the corresponding
encounter contribution must not be
added again as an independent loss term.

\subsection{Evolving networks and molecular relaxation}

In liquid water, the effective transfer
and termination rates depend on the
instantaneous hydrogen-bond configuration
${\cal C}(t)$. More generally,
\begin{equation}
w_{ij}=w_{ij}[{\cal C}(t);T],
\qquad
k_{s,a}=k_{s,a}[{\cal C}(t);T].
\label{eq:three_dynamic_rates}
\end{equation}

The complete stochastic problem must evolve
the network configuration and excitation
position jointly, using the microscopic
transition rates introduced in
Sec.~\ref{sec:network}. The network
cannot be treated as an independently
fluctuating background when excitation
transfer itself reconstructs hydrogen bonds.

The corresponding survival and first-passage
statistics may be nonexponential even if
the full configuration dynamics is Markovian.
Both the spatial organization and motion of
five-coordinated defects are relevant
to encounter annihilation.

The single-excitation calculation follows
an existing $3'$ defect until termination.
In equilibrium, new excitations are created
continually. Their creation and disappearance
must satisfy detailed balance through the
underlying reversible network transitions.
For the stationary active-defect population,
the mean creation flux equals the combined
mean flux through the two termination
channels, provided the classification
includes all relevant events.

Finally, the number of defect hops cannot
be identified directly with a molecular
relaxation time. Each transfer produces
a particular distribution of molecular
reorientations and displacements, and
successive events may be correlated.
The equilibrium population $c_{3'}(T)$,
the propagation statistics, and these
molecular changes must be combined to
calculate orientational relaxation and
transport.

We therefore first determine the survival,
annihilation probabilities, lifetimes,
and propagation activity of the active
$3'$ excitation. Their numerical evaluation,
beginning with the homogeneous approximation
and proceeding to network-resolved dynamics,
is presented in Sec.~VI.

%============================================
% =============== SECTION 4 ================

\section{Defect-Controlled Relaxation:
Mean-Field Theory and Correlation Effects}
\label{sec:defect_exponent}

We now turn to the central problem of this paper:
the relationship between the concentration of
active defects and the structural relaxation time.
We first derive the result for independently
propagating defects within a homogeneous mean-field
approximation. We then examine how the propagation
rate, finite defect lifetime, and correlations
among successive rearrangements modify this result.
This provides a general expression for the
effective relaxation exponent and identifies
the possible origins of its noninteger value.

\subsection{Definition of the defect exponent}

We now examine how the equilibrium concentration
of active $3'$ defects controls molecular
relaxation. The microscopic picture developed
in the preceding sections identifies relaxation
with the propagation of coordination defects
through the hydrogen-bond network. Each
successful transfer can induce molecular
reorientation and reconstruction of the
surrounding network.

Let $\tau_\alpha(T)$ denote the characteristic
structural relaxation time and $c_{3'}(T)$
the equilibrium concentration of active
three-coordinated defects. We introduce
the effective exponent $\alpha_D$ through
\begin{equation}
\tau_\alpha(T)\sim
[c_{3'}(T)]^{-\alpha_D}.
\label{eq:alpha_scaling}
\end{equation}

Since both quantities depend on temperature,
the local exponent is defined by
\begin{equation}
\alpha_D(T)=
-\frac{d\ln\tau_\alpha(T)}
{d\ln c_{3'}(T)}.
\label{eq:alpha_definition}
\end{equation}

This definition does not assume that
$\alpha_D$ is universal or asymptotically
constant. In particular, the steep
concentration dependence observed in
the simulations of Saito, Bagchi, and
Ohmine \cite{Saito2018} may represent an
effective exponent over a finite
temperature interval.

\subsection{Independent-defect mean-field theory}

We first consider a homogeneous population
of independently propagating $3'$ defects.

Let $k_0$ denote the effective rate at
which one active defect generates
relaxation events for a representative
molecule, including the geometrical
normalization associated with the number
of molecular sites. We assume that
$k_0$ is independent of defect
concentration.

In the mean-field approximation,
the relaxation events are statistically
independent and uniformly distributed.
The probability $C(t)$ that a local
configuration has retained its
initial memory obeys
\begin{equation}
\frac{dC(t)}{dt}
=-k_0c_{3'}C(t).
\label{eq:MF_relaxation}
\end{equation}

With $C(0)=1$, the solution is
\begin{equation}
C(t)=\exp[-k_0c_{3'}t].
\label{eq:MF_correlation}
\end{equation}

Thus,
\begin{equation}
\tau_\alpha^{\rm MF}
=\frac{1}{k_0c_{3'}}.
\label{eq:MF_tau}
\end{equation}

At fixed $k_0$, Eqs.~(\ref{eq:alpha_definition})
and (\ref{eq:MF_tau}) give
\begin{equation}
\alpha_D^{\rm MF}=1.
\label{eq:MF_alpha}
\end{equation}

This is the fundamental mean-field
prediction. The relaxation rate is
proportional to the number of
independently acting mobile defects.

The result does not require the defects
to have infinite lifetimes. Their
creation and disappearance maintain
the stationary concentration $c_{3'}$,
while the microscopic termination
processes influence the value of
$k_0$. Equation~(\ref{eq:MF_alpha})
requires that the effective
relaxation rate per defect remain
independent of $c_{3'}$.

\subsection{Network correlations and
relaxation efficiency}

The independent-defect approximation
neglects the correlations generated
by the hydrogen-bond network.

A propagating defect may revisit the
same molecular environment, induce
correlated reorientations, or encounter
regions in which further propagation
is inhibited. Moreover, the trajectory
is terminated by self-decay or
annihilation with a five-coordinated
defect, as described in Sec.~III.

To include these effects, we write
the effective relaxation rate as
\begin{equation}
k_{\rm rel}(T)=k_h(T)F(T),
\label{eq:correlation_rate}
\end{equation}
where $k_h$ is the mean rate of
successful defect-propagation events,
with the appropriate site
normalization, and $F$ is a
dimensionless relaxation-efficiency
factor.

The factor $F$ accounts for the
extent to which propagation events
produce independent, effective
molecular relaxation. It includes
the influence of repeated visits,
network correlations, and the
statistics of terminating
trajectories.

Assuming that productive defect-mediated rearrangements generate an approximately exponential decay of structural memory, the stationary relaxation rate can be written as

\begin{equation}
\tau_\alpha^{-1}
=c_{3'}k_hF.
\label{eq:correlated_tau}
\end{equation}

Equation (72) may be understood as a kinetic
counting relation rather than an independently
postulated power law. At equilibrium, the number
of active defects per molecule is $c_{3'}$, and
each active defect executes successful propagation
events at an average rate $k_h$. Their product
therefore measures the rate of defect-mediated
rearrangements per molecule. However, not every
propagation event produces an independent loss
of structural memory. Repeated visits to the same
environment, correlated molecular reorientations,
and overlap between trajectories reduce the
effective relaxation produced by a given number
of hops. These effects are represented by the
dimensionless factor $F$. In the independent-event
limit, where each successful hop contributes one
statistically independent relaxation event with
the adopted normalization, $F=1$. More generally,
$F$ must be determined from the propagation
statistics and the associated molecular
reorientations. Thus Eq.~(72) constitutes a
stationary effective-rate approximation, whose
microscopic content resides in the calculation
of $k_h$ and $F$, rather than an assumed
power-law dependence on defect concentration.

The independent-event limit has
$F=1$ with the corresponding
normalization of $k_h$.

An interpretation of $F$ follows
from the trajectories of the mobile
defects. Let $\langle n_{\rm hop}\rangle$
denote the mean number of successful
propagation events before termination,
and $\langle n_{\rm new}\rangle$ the
mean number of distinct molecular
environments productively relaxed
by those events.

For an idealized renewal process in
which a first productive visit
fully relaxes the local environment,
we may define
\begin{equation}
F_{\rm visit}=
\frac{\langle n_{\rm new}\rangle}
{\langle n_{\rm hop}\rangle}.
\label{eq:visit_efficiency}
\end{equation}

This quantity measures the loss of
efficiency caused by repeated visits
during individual excitation
trajectories. A more general
relaxation factor must also account
for partial molecular reorientation
and overlap between different
defect trajectories.

Importantly, the finite defect
lifetime must not be multiplied
independently into
Eq.~(\ref{eq:correlated_tau}).
Its effects already enter the
stationary population and
the trajectory statistics
used to determine $F$.

\subsection{Origin of the nontrivial exponent}

Taking the logarithmic derivative of
Eq.~(\ref{eq:correlated_tau}), we obtain
\begin{equation}
\alpha_D(T)=
1+
\frac{d\ln k_h}{d\ln c_{3'}}
+
\frac{d\ln F}{d\ln c_{3'}}.
\label{eq:alpha_general}
\end{equation}

This equation separates the
mean-field contribution from
two sources of additional
temperature dependence:
the elementary propagation
rate and the relaxation
efficiency arising from
network correlations.

If the product $k_hF$
decreases upon cooling
together with $c_{3'}$,
then
\begin{equation}
\alpha_D>1.
\label{eq:alpha_greater}
\end{equation}

The converse is also possible.
If the remaining active defects
become more efficient at exploring
the network, the correlation
contribution may reduce the
effective exponent.

Thus a large exponent is not
a universal consequence of
network connectivity. It reflects
the competition between defect
depletion, propagation, and
the effectiveness of the
resulting rearrangements.

A useful analytical illustration
is obtained by assuming
approximately activated behavior
over a restricted temperature range:
\begin{equation}
c_{3'}(T)\simeq c_0
\exp\left(-\frac{E_f}{k_BT}\right),
\label{eq:formation_energy}
\end{equation}
and
\begin{equation}
k_h(T)F(T)\simeq k_{\mathrm{act}}^{(0)}
\exp\left(-\frac{E_m}{k_BT}\right).
\label{eq:migration_energy}
\end{equation}

Here $c_0$ is a dimensionless concentration
prefactor, while $k_{\mathrm{act}}^{(0)}$
is a rate prefactor. The quantities
$E_f$ and $E_m$ characterize, respectively,
the temperature dependence of the active-defect
population and the productive relaxation rate
per defect, including propagation and
correlation effects.

Eliminating temperature gives
\begin{equation}
\alpha_D=
1+\frac{E_m}{E_f}.
\label{eq:alpha_energies}
\end{equation}

This result demonstrates how a
noninteger exponent can arise
without requiring a prescribed
number of cooperative defects.

For non-Arrhenius behavior,
we introduce local apparent
activation energies
\begin{equation}
E_f^{\rm app}(T)=
-k_B\frac{d\ln c_{3'}}{d(1/T)},
\label{eq:Ef_apparent}
\end{equation}
and
\begin{equation}
E_m^{\rm app}(T)=
-k_B\frac{d\ln(k_hF)}{d(1/T)}.
\label{eq:Em_apparent}
\end{equation}

The corresponding result is
\begin{equation}
\alpha_D(T)=
1+\frac{E_m^{\rm app}(T)}
{E_f^{\rm app}(T)}.
\label{eq:alpha_apparent}
\end{equation}

Equation~(\ref{eq:alpha_apparent})
does not require Arrhenius
temperature dependence and
makes explicit that the
observed exponent may vary
continuously with temperature.

\subsection{Microscopic interpretation
and connection to thermodynamics}

Equations~(\ref{eq:alpha_general})
and (\ref{eq:alpha_apparent})
provide a framework for
understanding the concentration
dependence of structural relaxation.

The mean-field exponent is unity.
Deviations arise because the
relaxation generated by each
active defect is itself
temperature dependent.

The two annihilation processes
influence this behavior through
the survival and exploration
statistics of defect trajectories.
Their contributions must be
calculated from the network
transition dynamics rather than
introduced as independent
multiplicative factors.

The quantity $F$ is not a
phenomenological power law
to be selected to reproduce
the simulations. Its
determination from the
network propagation operator,
including repeated visits
and correlated bond exchanges,
is a central task of the
quantitative theory.

A remaining requirement is the
equilibrium concentration
$c_{3'}(T,P)$ itself.

The thermodynamic theory
developed previously
\cite{BagchiDefect2026}
determines the collective
defective-environment
concentration $c_d(T,P)$.
However, $c_{3'}$ represents
a particular donor--acceptor
coordination state and cannot
be identified directly with
$c_d$.

The statistical relation
between these concentrations,
their temperature dependence,
and the corresponding
coordination-state
correlations will be examined
in the following section.

%================= Section 5 ===============
% ================== 5 ===================
\section{Microscopic Propagation and Statistical
Mechanics of Active Coordination Defects}
\label{sec:microscopic_defects}

The preceding section establishes the relation
between defect concentration and relaxation,
but leaves the microscopic propagation rate
and the equilibrium populations of active defects
to be determined. We now address these two
questions. We propose a concerted three-molecule
hydrogen-bond exchange mechanism, derive its
effective propagation rate, and examine how
the population of active coordination defects
can be connected to the statistical thermodynamics
of the hydrogen-bond network.

\subsection{Hydrogen-bond exchange and the
origin of defect mobility}

The molecular mechanism of water reorientation
was clarified by Laage and Hynes
\cite{Laage2006}, who demonstrated that
large-amplitude rotational jumps arise
through hydrogen-bond partner exchange,
rather than through a succession of
small diffusive rotations.

Subsequently, Jana, Singh, and Bagchi
\cite{Jana2011} demonstrated that
five-coordinated defects can propagate
through supercooled water in a string-like
fashion, producing successive molecular
rotational jumps.

Saito, Bagchi, and Ohmine
\cite{Saito2018} identified a particularly
important role of specific three-coordinated
defects in maintaining structural relaxation
in deeply supercooled water, even after
large defect-rich clusters have fragmented.

These results suggest a connection between
local hydrogen-bond exchange, the propagation
of coordination defects, and collective
structural relaxation.

We now examine a possible microscopic
mechanism for the propagation of the active
three-coordinated species
$3'=H^1O^2$.

This molecule possesses one unsatisfied
OH donor, while retaining two accepted
hydrogen bonds. A suitably oriented
intermolecular vibration can bring this
donor into a configuration favorable
for forming a new hydrogen bond.

Consider three molecules $i,j,k$,
initially having coordination states
$3',4,4$, respectively. Suppose that
molecule $k$ donates a hydrogen bond
to molecule $j$.

Formation of an additional hydrogen
bond from $i$ to $j$ produces
\begin{equation}
3'_i+4_j+4_k
\longrightarrow
4_i+5_j+4_k.
\label{eq:five_intermediate}
\end{equation}

The intermediate molecule $j$ is
temporarily five-coordinated, having
two donated and three accepted
hydrogen bonds. The coordination labels in Eq. (82) specify the hydrogen-bond states of the three participating molecules, without assigning the four-coordinated molecules to either ideal or defective collective environments. Such environmental distinctions enter the statistical counting of eligible configurations discussed in Sec. 5.5.

The additional accepted bond may
destabilize the original bond
from $k$ to $j$. Its subsequent
breaking produces
\begin{equation}
4_i+5_j+4_k
\longrightarrow
4_i+4_j+3'_k.
\label{eq:three_transfer_second}
\end{equation}

The net process is therefore
\begin{equation}
3'_i+4_j+4_k
\longrightarrow
4_i+4_j+3'_k.
\label{eq:three_transfer_net}
\end{equation}

The coordination labels in Eqs. (82)–(84) specify the hydrogen-bond states of the participating molecules but do not distinguish between ideal and defective collective environments. The exchange requires a suitable donor–acceptor arrangement: the initially unsatisfied donor of molecule (i) must be able to form a hydrogen bond with molecule (j), while an existing hydrogen bond donated by molecule (k) to (j) is subsequently broken. Thus, the coordination sequence specifies a possible microscopic mechanism, whereas its statistical occurrence depends on the local network geometry and connectivity. The active defect has propagated from molecule $i$ to molecule $k$
through a transient five-coordinated
intermediate at $j$.

The number of hydrogen bonds is
the same in the initial and final
configurations. During the intermediate
stage, one additional hydrogen bond
is temporarily present.

This mechanism satisfies donor--acceptor
bond bookkeeping and represents a
specific realization of the effective
defect transfer introduced in Sec.~III.
It requires suitable local geometry
and remains a microscopic hypothesis
whose relative importance must be
tested against other exchange pathways.
%=================================

\subsection{Why the $3'$ defect is dynamically active}

The proposed exchange mechanism provides a microscopic
reason for the dynamical distinction between the two
principal three-coordinated states. The active species
$3'=H^1O^2$ possesses one unsatisfied donor hydrogen
while retaining two accepted hydrogen bonds. This
unsatisfied donor can participate in the formation of
an additional hydrogen bond with a neighboring
four-coordinated molecule, producing the transient
five-coordinated intermediate of Eq.~(82).

The subsequent breaking of an existing hydrogen bond
accepted by the intermediate molecule transfers the
three-coordinated defect to another site. Thus, defect
propagation involves two successive bond rearrangements:
formation of a new hydrogen bond and breaking of an
existing one. The coordination mechanism conserves the
number of hydrogen bonds between the initial and final
configurations.

The alternative three-coordinated species $H^2O^1$
has two donated hydrogen bonds and only one accepted
bond. It does not possess the same unsatisfied donor
and therefore cannot initiate this particular
donor-addition mechanism without rearranging an
existing bond. This distinction does not exclude
other possible propagation mechanisms involving
$H^2O^1$.

An important feature of the proposed mechanism is
its dependence on the local network geometry.
A successful exchange requires the unsatisfied
donor of the active defect to approach a suitable
acceptor and an existing donor--acceptor bond to
become susceptible to breaking. Coordination
numbers alone do not determine whether these
conditions are satisfied.

The vibrational properties reported by Saito,
Bagchi, and Ohmine~\cite{Saito2018}
are relevant in this context. The enhanced
vibrational contribution in the region of
approximately $300$--$400\,\mathrm{cm}^{-1}$
suggests a possible coupling between local
coordination defects and intermolecular network
motions. Such motions may assist the formation
and subsequent decay of the transient
five-coordinated intermediate.

The dynamical activity of $3'$ therefore reflects
both its particular hydrogen-bond coordination
and the availability of favorable local exchange
configurations. The former identifies the active
species, while the latter determines the frequency
of successful propagation events.
%===================================

\subsection{Derivation of the effective
propagation rate}

We next derive the effective rate
of the two-step process. This
also provides a simple illustration
of the elimination of a short-lived
reaction intermediate.

For a specified eligible
three-molecule configuration,
we denote the initial state by
$A$, the five-coordinated
intermediate by $I$, and the
final transferred-defect state
by $B$.

The elementary kinetic scheme is
\begin{equation}
A
\mathop{\rightleftharpoons}^{k_1}_{k_{-1}}
I
\xrightarrow{k_2}
B.
\label{eq:elementary_kinetics}
\end{equation}

Here $k_1$ is the rate of
formation of the additional
hydrogen bond, $k_{-1}$
is the rate of its reversal,
and $k_2$ is the rate at
which the old hydrogen bond
breaks and transfers the defect.

The quantities $k_1$, $k_{-1}$,
and $k_2$ are first-order
transition rates for a specified
local configuration, each having
dimensions of inverse time.
They are not bulk chemical
rate constants.

Let $P_A(t)$, $P_I(t)$,
and $P_B(t)$ denote the
probabilities of occupying
the three configurations.

The kinetic equations are
\begin{equation}
\frac{dP_A}{dt}
=-k_1P_A+k_{-1}P_I,
\label{eq:kinetic_A}
\end{equation}
\begin{equation}
\frac{dP_I}{dt}
=k_1P_A-(k_{-1}+k_2)P_I,
\label{eq:kinetic_I}
\end{equation}
and
\begin{equation}
\frac{dP_B}{dt}=k_2P_I.
\label{eq:kinetic_B}
\end{equation}

For this first-passage calculation,
$B$ is treated as absorbing.
The reverse transfer can be
included when constructing the
full equilibrium network dynamics.

Suppose the intermediate relaxes
much faster than the initial
population changes appreciably:
\begin{equation}
k_{-1}+k_2\gg k_1.
\label{eq:intermediate_condition}
\end{equation}

After a short transient,
the intermediate population
is approximately stationary,
\begin{equation}
\frac{dP_I}{dt}\simeq0.
\label{eq:steady_intermediate}
\end{equation}

Equation~(\ref{eq:kinetic_I})
then gives
\begin{equation}
P_I\simeq
\frac{k_1}{k_{-1}+k_2}P_A.
\label{eq:intermediate_population}
\end{equation}

Substituting into the equation
for product formation,
\begin{equation}
\frac{dP_B}{dt}
\simeq
\frac{k_1k_2}{k_{-1}+k_2}P_A.
\label{eq:product_rate}
\end{equation}

Thus, the effective propagation
rate is
\begin{equation}
k_h^{\rm eff}
\simeq
\frac{k_1k_2}{k_{-1}+k_2}.
\label{eq:effective_hopping}
\end{equation}

This result has a transparent
physical interpretation.

Once the intermediate $I$
is formed, it can either
return to $A$ or proceed
to $B$. The probability
of proceeding to $B$ before
returning to $A$ is
\begin{equation}
p_{\rm trans}
=\frac{k_2}{k_{-1}+k_2}.
\label{eq:transmission_probability}
\end{equation}

Therefore,
\begin{equation}
k_h^{\rm eff}\simeq
k_1p_{\rm trans}.
\label{eq:attempt_transmission}
\end{equation}

The effective propagation
rate is the rate of forming
the intermediate multiplied
by the probability that
the intermediate completes
the transfer rather than
returning to the initial state.

For completeness, the exact
mean first-passage time
from $A$ to $B$ follows
from the same kinetic scheme.

Let $T_A$ and $T_I$ denote
the corresponding mean
remaining first-passage times.
Conditioning on the next
transition gives
\begin{equation}
T_A=\frac{1}{k_1}+T_I,
\label{eq:MFPT_A}
\end{equation}
and
\begin{equation}
T_I=
\frac{1}{k_{-1}+k_2}
+\frac{k_{-1}}{k_{-1}+k_2}T_A.
\label{eq:MFPT_I}
\end{equation}

Solving these two equations,
we obtain
\begin{equation}
T_A=
\frac{k_1+k_{-1}+k_2}{k_1k_2}.
\label{eq:exact_MFPT}
\end{equation}

The inverse mean first-passage
time is consequently
\begin{equation}
\frac{1}{T_A}
=
\frac{k_1k_2}
{k_1+k_{-1}+k_2}.
\label{eq:exact_inverse_MFPT}
\end{equation}

In the short-lived-intermediate
limit of Eq.~(\ref{eq:intermediate_condition}),
this reduces to
Eq.~(\ref{eq:effective_hopping}).

Equation~(99) gives the inverse mean first-passage
time for completion of a specified exchange pathway,
starting from the initial configuration $A$.
It retains the residence time in the transient
intermediate and does not require the
short-lived-intermediate approximation of
Eq.~(89). In a coarse-grained description where
completed transfers are treated as renewal events,
this inverse mean first-passage time provides a
natural effective propagation rate. The
corresponding exponential relaxation rate of the
three-state kinetic system need not be identical
to the inverse mean first-passage time.

The distinction is useful:
the effective hopping expression
is an adiabatic elimination
of the intermediate, whereas
Eq.~(\ref{eq:exact_MFPT})
retains the time spent in
both the initial and
intermediate configurations.
%===============================

\subsection{Coordination statistics and
the population of active defects}

The microscopic propagation
mechanism must now be connected
to equilibrium statistical mechanics.

Let
\begin{equation}
p_{mn}(T,P)
=
\left\langle
\delta_{n_i^D,m}
\delta_{n_i^A,n}
\right\rangle
\label{eq:coordination_distribution}
\end{equation}
denote the equilibrium probability
of a molecule having $m$ donated
and $n$ accepted hydrogen bonds.

The average is taken over the equilibrium
network ensemble, with $i$ denoting a
representative water molecule.

The active-defect concentration is
\begin{equation}
c_{3'}(T,P)=p_{12}(T,P),
\label{eq:active_population}
\end{equation}
while the five-coordinated
population of interest is
\begin{equation}
c_5(T,P)=p_{23}(T,P).
\label{eq:five_population}
\end{equation}

The coordination probabilities
satisfy
\begin{equation}
\sum_{m,n}p_{mn}=1,
\label{eq:coord_normalization}
\end{equation}
and, from donor--acceptor
bond counting,
\begin{equation}
\sum_{m,n}(m-n)p_{mn}=0.
\label{eq:coord_balance}
\end{equation}

However, these constraints alone
are insufficient to determine
the individual populations.
Their values depend on local
energies, hydrogen-bond geometry,
and correlations between
neighboring molecular states.

The thermodynamic field theory
developed previously
\cite{BagchiDefect2026}
introduces a collective defect
concentration $c_d(T,P)$.
An ideal environment is one
in which the central molecule
and all its hydrogen-bonded
neighbors are four-coordinated.

This definition differs from
the molecular coordination
classification used here.

Since every $3'$ molecule
belongs to a defective
environment, the exact
statistical relation is
\begin{equation}
c_{3'}(T,P)
=
c_d(T,P)
P(3'|d;T,P).
\label{eq:conditional_active}
\end{equation}

The conditional probability
$P(3'|d)$ describes the
fraction of defective
environments whose central
molecule is specifically
$H^1O^2$.

It is not fixed by the
collective defect concentration
alone.

In the existing thermodynamic
theory, the reduced defect
free energy has the form
\begin{align}
f_d(c_d)={}&
k_BT\left[
c_d\ln c_d+
(1-c_d)\ln(1-c_d)
-c_d\ln g_d
\right]
\nonumber\\
&+\epsilon_d c_d
-J(1-c_d)^2.
\label{eq:collective_defect_free_energy}
\end{align}

The equilibrium collective
defect fraction follows from
minimization of the appropriate
thermodynamic potential, with
volume coupling included when
required.

Equation~(\ref{eq:collective_defect_free_energy})
does not by itself distinguish
the $H^1O^2$, $H^2O^1$,
and $H^2O^3$ populations.

To resolve these states
consistently, we must consider
the conditional distribution
of microscopic hydrogen-bond
configurations within a given
collective defect sector.

Schematically, if ${\cal C}$
denotes a microscopic network
configuration, the constrained
partition function is
\begin{equation}
Z(c_d)=
\sum_{\mathcal C}
e^{-\beta{\cal H}(\mathcal C)}
\delta_{c_d(\mathcal C),c_d}.
\label{eq:constrained_partition}
\end{equation}

Here $\mathcal H(C)$ denotes the microscopic
energy of the network configuration $C$,
including the energetic contributions associated
with the underlying molecular configurations,
and $\beta=1/(k_B T)$.

For a discrete finite system,
the constraint is understood
at an allowed value of $c_d$;
a narrow concentration interval
may be used in the
thermodynamic limit.

The corresponding conditional
active-defect population is
\begin{equation}
c_{3'}(c_d)=
\frac{1}{NZ(c_d)}
\sum_{\mathcal C}
N_{3'}(\mathcal C)
e^{-\beta{\cal H}(\mathcal C)}
\delta_{c_d(\mathcal C),c_d},
\label{eq:conditional_partition}
\end{equation}
where $N_{3'}(\mathcal C)$
is the number of
$H^1O^2$ molecules.

This formulation clarifies
the statistical-mechanical
problem. The microscopic
ensemble must reproduce the
collective defect thermodynamics
upon coarse-graining, while
also predicting the
coordination-resolved
populations.

The configurational entropy
already included in
Eq.~(\ref{eq:collective_defect_free_energy})
must not be counted again
as an independent ideal
mixing entropy of the
coordination species.
%===================================

\subsection{Network statistics and
the propagation rate}

The effective rate derived above
applies to one eligible
three-molecule configuration.

To obtain the average
propagation rate of an active
defect, we must count the
available configurations
and average over their
local transition rates.

Let $X_{ijk}(\mathcal C)$ be an indicator equal
to unity when the molecules $i,j,k$ form an
eligible directed exchange configuration in
the instantaneous hydrogen-bond network
$\mathcal C$, and zero otherwise.
Molecule $i$ carries the active $3'$ defect,
molecule $j$ is initially four-coordinated,
and molecule $k$ is initially four-coordinated
and donates a hydrogen bond to $j$.
In addition, the unsatisfied donor of $i$
must be able to form the additional hydrogen
bond required by Eq.~(82).

The equilibrium propagation rate per active
defect is then
\begin{equation}
k_h(T,P)=
\frac{
\left\langle
\sum_{ijk}
X_{ijk}(\mathcal C)
k_{h,ijk}^{\mathrm{eff}}(\mathcal C)
\right\rangle
}{
\langle N_{3'}\rangle
}.
\tag{109}
\end{equation}

Here $k_{h,ijk}^{\mathrm{eff}}$ is the effective
rate associated with the specified local
exchange pathway. The inverse mean first-passage
time derived in Eq.~(99) provides one possible
coarse-grained definition of this rate.

Equation~(109) separates the statistical
availability of exchange configurations from
their microscopic dynamics. The statistical
factor depends on the coordination states,
donor--acceptor connectivity, and local
geometry of the hydrogen-bond network.
It cannot, in general, be expressed as a
simple product of bulk coordination
concentrations.

In particular, the four-coordinated molecules
participating in Eq.~(82) need not belong
to different collective environments.
The collective defective-environment fraction
$c_d$ is distinct from the molecular
coordination populations. Consequently,
the product
$(c_d-c_{3'}-c_3-c_5)(1-c_d)$
would describe a restricted environmental
assignment rather than the complete set
of eligible exchange configurations.

A minimal network approximation may be
constructed by treating the hydrogen-bond
connectivity explicitly and counting the
allowed directed exchange pathways. A
four-coordinated acceptor has two incoming
hydrogen bonds and hence two possible
donor bonds whose breaking could complete
the exchange. These represent potential
channels, not necessarily independent
successful events. Their actual contribution
depends on the accessibility of the
intermediate and on the competition between
the possible bond-breaking processes.

The transient five-coordinated intermediate
in Eq.~(82) must also be distinguished from
the equilibrium population $c_5$.
The latter enters the equilibrium network
statistics and defect-termination processes,
but is not automatically the population of
the short-lived intermediate involved in
an individual exchange.

A quantitative evaluation of Eq.~(109)
therefore requires either coordination-resolved
network statistics or a specified microscopic
network model. This distinction is important
when connecting the equilibrium defect
thermodynamics to the propagation rate
and ultimately to structural relaxation.

Finally, combining the
microscopic propagation rate
with the relaxation
efficiency $F(T)$ introduced
in Sec.~IV gives
\begin{equation}
\tau_\alpha^{-1}(T)
\simeq
c_{3'}(T)k_h(T)F(T).
\label{eq:microscopic_relaxation_rate}
\tag{109a}
\end{equation}

Consequently,
\begin{equation}
\alpha_D(T)=
1+
\frac{d\ln k_h}{d\ln c_{3'}}
+
\frac{d\ln F}{d\ln c_{3'}}.
\label{eq:microscopic_alpha}
\end{equation}

The mean-field result
$\alpha_D=1$ is recovered
when the productive
relaxation rate per
active defect is
concentration independent.

The nontrivial exponent
therefore reflects both
the equilibrium scarcity
of active defects and
the temperature dependence
of their propagation
and network-relaxation
efficiency.

The principal unresolved
statistical-mechanical task
is to determine the
coordination-resolved
conditional probabilities
from the collective defect
thermodynamics.

In the following section,
we examine the quantitative
consequences of the
analytical expressions
using a minimal network
model.

%================= Section 6 ============

\section{Defect Interactions and the
Relaxation Exponent}
\label{sec:results}

We now examine the quantitative consequences
of the preceding theory using a minimal model
of defect propagation on an idealized network.
Our objective is to determine how interactions
between defects, repeated visits to the same
molecular environment, and finite defect lifetimes
affect the relaxation exponent. We first develop
an analytical treatment for a four-connected
network and then present illustrative numerical
results. The calculation is intended to establish
the physical trends rather than provide a
quantitative fit to supercooled water.

\subsection{Theory}

We now examine how interactions between
mobile coordination defects modify the
mean-field exponent $\alpha_D=1$
derived in Sec.~IV.

Let $c=c_{3'}$ and $f=c_5$ denote
the equilibrium concentrations of
active three-coordinated and
five-coordinated defects, respectively.
A mobile $3'$ defect propagates through
hydrogen-bond exchange and can disappear
through intrinsic self-decay or
annihilation following an encounter
with a five-coordinated defect.

In a homogeneous approximation,
the total termination rate is
\begin{equation}
\mu=k_s+k_af,
\label{eq:VI_mu}
\end{equation}
Here $k_s$ is the intrinsic self-decay rate,
while $k_a$ is the effective encounter-rate
coefficient. The product $k_a f$ is the
mean-field annihilation hazard associated
with five-coordinated defects. Thus $\mu$
is the total termination rate of an
active excitation.

To represent the simplest effect
of interacting defects, we assume
that the propagation rate depends
on the local concentration of
active excitations:
\begin{equation}
k_h(c)=k_0(1+gc).
\label{eq:VI_facilitation}
\end{equation}
Here $k_0$ is the propagation rate of an
isolated active defect in the dilute limit,
and $g$ is a dimensionless facilitation
parameter. Both $k_h$ and $k_0$ have
dimensions of inverse time. This is a
phenomenological pair-level approximation,
rather than a microscopic result.

We first consider a mobile defect
performing a random walk on a
four-connected network. For
analytical simplicity, the network
is approximated by an infinite
regular tree.

Let $a$ be the probability that
a walker reaches a specified
neighboring site before
termination. First-step analysis
gives
\begin{equation}
a=\frac{1+3a^2}
{4(1+\mu/k_h)}.
\label{eq:VI_first_passage}
\end{equation}

Defining $x=\mu/k_h$, the physical
solution is
\begin{equation}
a(x)=
\frac{2}
{4(1+x)+
\sqrt{16(1+x)^2-12}}.
\label{eq:VI_a}
\end{equation}

The average number of propagation
steps before termination is
\begin{equation}
\langle n_{\rm hop}\rangle
=\frac{k_h}{\mu}.
\label{eq:VI_hops}
\end{equation}

The corresponding average number
of distinct sites visited,
including the initial site, is
\begin{equation}
\langle N_{\rm distinct}\rangle
=1+\frac{4a}{1-3a}.
\label{eq:VI_distinct}
\end{equation}

We define the relaxation efficiency
as the fraction of propagation
steps reaching previously unvisited
sites:
\begin{equation}
F(x)=\frac{4xa(x)}{1-3a(x)}.
\label{eq:VI_F}
\end{equation}

Within the independent-visit
relaxation approximation,
\begin{equation}
\tau_\alpha^{-1}=c\,k_h(c)F(x).
\label{eq:VI_tau}
\end{equation}

The effective exponent is therefore
\begin{align}
\alpha_D
={}&1+\frac{gc}{1+gc}
\nonumber\\
&+
\frac{d\ln F}{d\ln x}
\left[
\frac{d\ln\mu}{d\ln c}
-\frac{gc}{1+gc}
\right].
\label{eq:VI_alpha}
\end{align}

The first term is the independent-defect
mean-field result. The second describes
facilitation of propagation, while the
last represents the influence of
repeated visits and finite defect
lifetimes.

For concentration-independent
microscopic rates and fixed network
statistics, the exponent reduces
to unity. A larger exponent requires
the productive relaxation activity
per defect to change with concentration.

\subsection{Numerical results}

We evaluate the preceding expressions
using illustrative dimensionless
parameters
\begin{equation}
k_0=1,\qquad g=10,\qquad
k_s=0.01,\qquad k_a=2,
\label{eq:VI_parameters}
\end{equation}
and assume
\begin{equation}
c_5=0.5c_{3'}.
\label{eq:VI_concentrations}
\end{equation}

The latter relation is adopted only
for this model calculation and is
not inferred from the SBO simulations.

\begin{table}[htbp]
\caption{
Illustrative results for interacting mobile
coordination defects on a four-connected tree.
The active three-coordinated defect concentration
is denoted by $c_{3'}$.
The calculations use the dimensionless parameters
$k_0=1$, $g=10$, $k_s=0.01$, and $k_a=2$,
with the assumed concentration relation
$c_5=0.5c_{3'}$.
Here $F$ is the relaxation-efficiency factor,
which accounts for repeated visits and the
finite lifetime of propagating defects;
$\tau_\alpha$ is the structural relaxation time,
expressed in inverse $k_0$ units;
and $\alpha_D$ is the effective exponent
describing the concentration dependence of
$\tau_\alpha$.
The results illustrate how concentration-dependent
defect propagation and reactive termination
can lead to an effective exponent larger than
the independent-defect mean-field value
$\alpha_D=1$.
These results are illustrative and are not
a quantitative fit to simulations of
supercooled water.
}
\label{tab:defect_results}

\centering

\begingroup
\large
\renewcommand{\arraystretch}{1.6}
\setlength{\tabcolsep}{18pt}

%\begin{ruledtabular}
\begin{tabular}{cccc}
$c_{3'}$ & $F$ & $\tau_\alpha$ & $\alpha_D$ \\
0.005 & 0.676 & 281.85 & 1.051 \\
0.010 & 0.678 & 134.05 & 1.098 \\
0.020 & 0.682 & 61.08  & 1.177 \\
0.050 & 0.691 & 19.31  & 1.349 \\
0.100 & 0.698 & 7.16   & 1.516 \\
0.200 & 0.706 & 2.36   & 1.680 \\
\end{tabular}
%\end{ruledtabular}

\endgroup
\end{table}

The results demonstrate that
interacting defect propagation
can increase the effective
relaxation exponent above its
mean-field value of unity.

The influence of encounter
annihilation enters through the
finite lifetime and the fraction
of productive propagation events.
In the present model, this
contribution is relatively modest,
whereas facilitation produces
the larger increase.

An important analytical feature
is that $F$ remains bounded:
\begin{equation}
\frac{2}{3}\leq F\leq1.
\label{eq:VI_F_bounds}
\end{equation}

Thus, repeated visits on this
idealized network modify the
relaxation efficiency but cannot,
by themselves, produce an
arbitrarily large exponent.

The calculated $\alpha_D$ is
not universal and approaches
unity in the dilute limit
for the present choice of rates.

The substantially stronger
concentration dependence reported
for deeply supercooled water
\cite{Saito2018} may involve
additional changes in network
structure, exchange kinetics,
and cooperative relaxation as
the liquid is cooled.

Our objective here is more
limited: to establish analytically
and numerically that correlated
defect propagation and reactive
encounters can produce departures
from the independent-defect
result $\alpha_D=1$.

A fully quantitative theory
would require temperature-dependent
coordination populations and
microscopic propagation rates,
consistent with the thermodynamic
description of Sec.~V.
%
%=================  Section 7 =========================

\section{Conclusions and Outlook}
\label{sec:conclusions}

Let me first summarize the main results of the paper.
Intrigued by the interplay between coordination defects
in water and the relaxation of its hydrogen-bonded network,
and particularly by the observed power-law dependence of
the structural relaxation time on the concentration of
active three-coordinated defects, we have developed a
microscopic defect-based theory of network relaxation.
The simulations of Saito, Bagchi, and Ohmine
\cite{Saito2018} revealed the remarkable persistence
of relaxation even when extended defect-rich regions
fragment upon deep supercooling. They also uncovered
an unusually strong dependence of the relaxation time
on the population of the $H^1O^2$ defect, denoted
here by $3'$. Our principal objective has been to
understand the physical origin of this behavior,
rather than to reproduce a particular numerical
exponent.

An important result is the identification of a
possible microscopic mechanism for the exceptional
activity of the $3'$ defect. The unsatisfied donor
hydrogen can form an additional hydrogen bond,
creating a transient five-coordinated intermediate.
The subsequent breaking of another hydrogen bond
transfers the three-coordinated defect to a different
molecule. This concerted three-molecule process
connects the molecular jump mechanism of Laage
and Hynes \cite{Laage2006,Laage2008} with the
propagation of coordination defects demonstrated
by Jana, Singh, and Bagchi \cite{Jana2011}.
The mechanism also provides a possible explanation
for the dynamical asymmetry between different
three-coordinated water molecules. We emphasize
that the proposed pathway requires favorable local
geometry and should be tested against other possible
hydrogen-bond exchange mechanisms. The important
role of coordination defects in network reconstruction
is also evident in studies of ice melting
\cite{Mochizuki2013,OhmineSaito2021}.

A second result concerns the relationship between
defect concentration and structural relaxation.
For a homogeneous population of independently
propagating defects, the relaxation time is inversely
proportional to the concentration of active defects,
giving the mean-field exponent $\alpha_D=1$.
We find that the effective exponent can increase
when interactions between mobile defects alter
their propagation rates and the efficiency with
which their trajectories produce molecular
rearrangements. The competition between propagation,
self-decay, and encounter annihilation with
five-coordinated defects provides an additional
source of dynamical correlations. Our analytical
expressions and simple network calculations
demonstrate how a noninteger exponent can arise
without assuming a prescribed number of
cooperatively rearranging molecules.

There is an interesting historical connection
here with the defect-diffusion description of
dielectric relaxation introduced by Glarum
\cite{Glarum1960,Glarum1960Polar} and developed
further by Bordewijk \cite{Bordewijk1975}.
The present problem is, however, more complicated.
The propagating excitation continually reconstructs
the network on which it moves, and its motion is
coupled to the creation, disappearance, and
interaction of other coordination defects.
The hydrogen-bond network is therefore both the
medium supporting defect propagation and a
dynamical object modified by that propagation.
This distinction is central to our approach.

The equilibrium statistical mechanics of defects
provides another essential ingredient. Our earlier
thermodynamic theory \cite{BagchiDefect2026}
describes collective defective environments
through their configurational entropy and the
cooperative stabilization of the tetrahedral network.
The concentration of such environments is not
identical to that of the dynamically active $3'$
species. Connecting the two requires
coordination-resolved statistical mechanics
consistent with the underlying network entropy.
This remains an important problem for further
investigation. It should ultimately allow the
equilibrium populations of active defects and
their temperature dependence to be calculated
within the same thermodynamic framework.

We should also recognize the limitations of the
present treatment. The numerical calculations
employ an idealized network and simplified
interaction kinetics. They establish the
possibility of an enhancement of the relaxation
exponent above its mean-field value, but do not
constitute a quantitative explanation of the
large effective exponent reported in deeply
supercooled water. Additional collective processes,
including changes in tetrahedral organization
and the availability of hydrogen-bond exchange
pathways, may become important upon strong
supercooling. The complex temperature dependence
of water dynamics has indeed been discussed in
several different theoretical and experimental
contexts \cite{Shi2018,Tyburski2026}.
We do not attempt to resolve all these phenomena
within the present model.

Finally, we return to the broader theoretical
motivation of this work. The conventional
mode-coupling description of liquid dynamics
emphasizes density fluctuations and their
self-consistent feedback. In network-forming
liquids, however, local bonding constraints,
coordination defects, and their propagation
introduce additional slow and spatially
correlated processes. The earlier studies
of bond-network lifetimes
\cite{Sciortino1990}, kinetically constrained
models \cite{Fredrickson1984}, and cooperative
elastic relaxation \cite{BagchiElastic2026}
illustrate complementary aspects of the
connection between local rearrangements
and collective slowing.

It is important to clarify the level of approximation
involved in the present theory. We have developed
a microscopic mechanism for the propagation of
active three-coordinated defects and formulated
their reaction--transport dynamics, including
propagation, self-decay, and annihilation through
encounters with five-coordinated defects. We have
also derived the effective propagation rate
associated with the transient five-coordinated
intermediate. The network calculations, however,
have been carried out within a simplified
statistical description, using an idealized
four-coordinated network and effective rates
for propagation and termination. The resulting
theory establishes the mean-field exponent
$\alpha_D=1$ and demonstrates how the
concentration dependence of defect propagation
and network-relaxation efficiency can give rise
to a larger, noninteger exponent. It does not
yet provide a quantitative explanation of the
large exponent observed in deeply supercooled
water \cite{Saito2018,OhmineSaito2021}.

A more complete description must treat the
hydrogen-bond network as an evolving,
correlated structure rather than merely
as the medium supporting defect motion.
Each propagation event changes local
hydrogen-bond connectivity and hence the
possible pathways for subsequent motion.
The creation, propagation, and annihilation
of defects are therefore dynamically coupled
to the reconstruction of the network itself.
A microscopic treatment should determine
these correlated processes, the statistics
of defect lifetimes and propagation lengths,
and the molecular reorientations generated
by successive bond exchanges. Equally
important is the connection between
equilibrium defect thermodynamics and the
populations of specific active coordination
states. The collective defect concentration
obtained from our thermodynamic theory
\cite{BagchiDefect2026} must ultimately be
resolved into the different donor--acceptor
species without double counting their
configurational entropy. These developments
are needed before a quantitative prediction
of the temperature dependence of structural
relaxation becomes possible.

Finally, we return to the broader theoretical
motivation of this work. The conventional
mode-coupling description of liquid dynamics
emphasizes density fluctuations and their
self-consistent feedback. In network-forming
liquids, however, local bonding constraints,
coordination defects, and their propagation
introduce additional dynamical processes
that may remain important even when the
population of defects becomes small.
The early studies of hydrogen-bond network
lifetimes \cite{Sciortino1990}, kinetically
constrained models \cite{Fredrickson1984},
and cooperative elastic relaxation
\cite{BagchiElastic2026} illustrate
complementary aspects of the connection
between local rearrangements and collective
slowing. The present description offers a
microscopic route for connecting these
different perspectives.

We envisage the present theory as a starting
point for a more complete mode-coupling-like
description in which defect populations,
their propagation and reaction dynamics,
and the evolving hydrogen-bond network
are coupled to conserved density and
momentum fluctuations. Such a formulation
would build upon the nonlinear fluctuating
hydrodynamic approach of Das and Mazenko
\cite{DasMazenko1986} and the treatment of
metastable defect variables by Yeo and
Mazenko \cite{YeoMazenko1995}, while replacing
phenomenological defect fields by physically
identifiable coordination excitations.
The retention of momentum fluctuations is
particularly important, because transverse
hydrodynamic modes provide relaxation
channels that must remain consistent with
momentum conservation. Such a theory could
also provide a useful perspective on the
dynamics of other network-forming liquids,
including silica, for which mode-coupling
descriptions have been developed
\cite{SciortinoKob2001}.

The broader message emerging from our study
is that structural relaxation in a network
liquid need not require the destruction
of the network itself. A relatively small
population of mobile coordination defects
can produce continuing molecular
rearrangements through successive
bond exchanges. The concentration of
these defects determines how many
relaxation-active excitations are present,
while their propagation and correlations
determine how effectively they relax
the surrounding network. Understanding
these two aspects together appears
essential for a microscopic theory of
relaxation in supercooled network liquids.

%====================================


\begin{thebibliography}{99}

\bibitem{Errington2001}
J. R. Errington and P. G. Debenedetti,
``Relationship between structural order and
the anomalies of liquid water,''
Nature \textbf{409}, 318--321 (2001).

\bibitem{Gibbs1973}
J. H. Gibbs, C. Cohen, P. D. Fleming,
and H. Porosoff,
``Toward a model for liquid water,''
J. Solution Chem. \textbf{2},
277--299 (1973).
doi:10.1007/BF00651977.

\bibitem{FlemingGibbs1974}
P. D. Fleming III and J. H. Gibbs,
``An adaptation of the lattice gas
to the water problem,''
J. Stat. Phys. \textbf{10},
157--173 (1974).
doi:10.1007/BF01009718.

\bibitem{FlemingGibbs1974II}
P. D. Fleming III and J. H. Gibbs,
``An adaptation of the lattice gas
to the water problem. II.
Second-order approximation,''
J. Stat. Phys. \textbf{10},
351--378 (1974).
doi:10.1007/BF01008799.

\bibitem{BagchiDonoghueGibbs1983}
B. Bagchi, E. Donoghue, and J. H. Gibbs,
``Agreement between the gelation and
molecular dynamics models of the
hydrogen-bond network in water,''
Chem. Phys. Lett. \textbf{94},
253--258 (1983).
doi:10.1016/0009-2614(83)87082-1.

\bibitem{Ito1999}
K. Ito, C. T. Moynihan, and C. A. Angell,
``Thermodynamic determination of fragility in liquids
and a fragile-to-strong liquid transition in water,''
Nature \textbf{398}, 492--495 (1999).
doi:10.1038/19072.

\bibitem{Bergman2000}
R. Bergman and J. Swenson,
``Dynamics of supercooled water in confined geometry,''
Nature \textbf{403}, 283--286 (2000).
doi:10.1038/35002027.

\bibitem{Mallamace2012}
F. Mallamace, C. Corsaro, P. Baglioni, E. Fratini,
and S.-H. Chen,
``The dynamical crossover phenomenon in bulk water,
confined water and protein hydration water,''
J. Phys.: Condens. Matter \textbf{24}, 064103 (2012).
doi:10.1088/0953-8984/24/6/064103.

\bibitem{Xu2016}
Y. Xu, N. G. Petrik, R. S. Smith, B. D. Kay,
and G. A. Kimmel,
``Growth rate of crystalline ice and the diffusivity
of supercooled water from 126 to 262 K,''
Proc. Natl. Acad. Sci. USA \textbf{113},
14921--14925 (2016).
doi:10.1073/pnas.1611395114.

\bibitem{Kringle2021}
L. Kringle, W. A. Thornley, B. D. Kay,
and G. A. Kimmel,
``Structural relaxation and crystallization in
supercooled water from 170 to 260 K,''
Proc. Natl. Acad. Sci. USA \textbf{118},
e2022884118 (2021).
doi:10.1073/pnas.2022884118.

\bibitem{Mussa2023}
A. Mussa, R. Berthelard, F. Caupin,
and B. Issenmann,
``Viscosity and Stokes--Einstein relation in deeply
supercooled water under pressure,''
J. Chem. Phys. \textbf{159}, 151103 (2023).
doi:10.1063/5.0169743.

\bibitem{Sciortino1990}
F. Sciortino, P. H. Poole, H. E. Stanley,
and S. Havlin,
``Lifetime of the bond network and gel-like
anomalies in supercooled water,''
Phys. Rev. Lett. \textbf{64}, 1686--1689 (1990).
doi:10.1103/PhysRevLett.64.1686.


\bibitem{Sciortino1991}
F. Sciortino, A. Geiger, and H. E. Stanley,
``Effect of defects on molecular mobility in
liquid water,''
Nature \textbf{354}, 218--221 (1991).
doi:10.1038/354218a0.

\bibitem{Sciortino1990Fifth}
F. Sciortino, A. Geiger, and H. E. Stanley,
``Isochoric differential scattering functions
in liquid water: The fifth neighbor as a
network defect,''
Phys. Rev. Lett. \textbf{65},
3452--3455 (1990).
doi:10.1103/PhysRevLett.65.3452.

\bibitem{Sciortino1992}
F. Sciortino, A. Geiger, and H. E. Stanley,
``Network defects and molecular mobility
in liquid water,''
J. Chem. Phys. \textbf{96},
3857--3865 (1992).

\bibitem{Laage2006}
D. Laage and J. T. Hynes,
``A molecular jump mechanism of water reorientation,''
Science \textbf{311}, 832--835 (2006).
doi:10.1126/science.1122154.

\bibitem{Jana2011}
B. Jana, R. S. Singh, and B. Bagchi,
``String-like propagation of the 5-coordinated
defect state in supercooled water: Molecular
origin of dynamic and thermodynamic anomalies,''
Phys. Chem. Chem. Phys. \textbf{13},
16220--16226 (2011).
doi:10.1039/C0CP02081H.

\bibitem{Dueby2019}
S. Dueby, V. Dubey, and S. Daschakraborty,
``Decoupling of translational diffusion from
the viscosity of supercooled water:
Role of translational jump diffusion,''
J. Phys. Chem. B \textbf{123},
7178--7189 (2019).
doi:10.1021/acs.jpcb.9b01719.

\bibitem{Dubey2021}
V. Dubey, S. Dueby, and S. Daschakraborty,
``Breakdown of the Stokes--Einstein relation
in supercooled water: The jump-diffusion
perspective,''
Phys. Chem. Chem. Phys. \textbf{23},
19964--19986 (2021).
doi:10.1039/D1CP02202D.

\bibitem{Nagaraj2026}
V. Nagaraj and S. Daschakraborty,
``Structural origin of translational jumps
and allied dynamical anomalies in
supercooled water,''
J. Phys. Chem. B \textbf{130},
2217--2225 (2026).
doi:10.1021/acs.jpcb.5c07141.


\bibitem{Saito2018}
S. Saito, B. Bagchi, and I. Ohmine,
``Crucial role of fragmented and isolated defects
in persistent relaxation of deeply supercooled
water,''
J. Chem. Phys. \textbf{149}, 124504 (2018).
doi:10.1063/1.5044458.

\bibitem{OhmineSaito2021}
I. Ohmine and S. Saito,
``Dynamical behavior of water; fluctuation,
reactions and phase transitions,''
Bull. Chem. Soc. Jpn. \textbf{94},
2575--2601 (2021).
doi:10.1246/bcsj.20210269.


\bibitem{BagchiDefect2026}
B. Bagchi,
``Defect Entropy and Cooperative Fluctuations
in Supercooled Liquid Water,''
 The Journal of Physical Chemistry, in Press (2026).

\bibitem{SciortinoKob2001}
F. Sciortino and W. Kob,
``Debye--Waller factor of liquid silica:
Theory and simulation,''
Phys. Rev. Lett. \textbf{86}, 648--651 (2001).
doi:10.1103/PhysRevLett.86.648.

\bibitem{DasMazenko1986}
S. P. Das and G. F. Mazenko,
``Fluctuating nonlinear hydrodynamics and the
liquid-glass transition,''
Phys. Rev. A \textbf{34}, 2265--2282 (1986).
doi:10.1103/PhysRevA.34.2265.

\bibitem{YeoMazenko1995}
J. Yeo and G. F. Mazenko,
``Metastable dynamics above the glass transition,''
Phys. Rev. E \textbf{51}, 5752--5761 (1995).
doi:10.1103/PhysRevE.51.5752.

\bibitem{Luzar1996}
A. Luzar and D. Chandler,
``Hydrogen-bond kinetics in liquid water,''
Nature \textbf{379}, 55--57 (1996).
doi:10.1038/379055a0.

\bibitem{Tyburski2026}
R. Tyburski, M. Shin, S. You, \textit{et al.},
``Observation of a dynamic transition in bulk
supercooled water,''
Nat. Phys. \textbf{22}, 21--26 (2026).
doi:10.1038/s41567-025-03112-3.

\bibitem{Shi2018}
R. Shi, J. Russo, and H. Tanaka,
``Origin of the emergent fragile-to-strong
transition in supercooled water,''
Proc. Natl. Acad. Sci. USA \textbf{115},
9444--9449 (2018).
doi:10.1073/pnas.1807821115.

\bibitem{Glarum1960}
S. H. Glarum,
``Dielectric relaxation of isoamyl bromide,''
J. Chem. Phys. \textbf{33}, 639--643 (1960).
doi:10.1063/1.1731229.

\bibitem{Glarum1960Polar}
S. H. Glarum,
``Dielectric relaxation of polar liquids,''
J. Chem. Phys. \textbf{33}, 1371--1375 (1960).
doi:10.1063/1.1731414.

\bibitem{Bordewijk1975}
P. Bordewijk,
``Defect-diffusion models of dielectric
relaxation,''
Chem. Phys. Lett. \textbf{32}, 592--596 (1975).
doi:10.1016/0009-2614(75)85248-1.

\bibitem{Laage2008}
D. Laage and J. T. Hynes,
``On the molecular mechanism of water reorientation,''
J. Phys. Chem. B \textbf{112},
14230--14242 (2008).
doi:10.1021/jp805217u.

\bibitem{Mochizuki2013}
K. Mochizuki, M. Matsumoto, and I. Ohmine,
``Defect pair separation as the controlling
step in homogeneous ice melting,''
Nature \textbf{498}, 350--354 (2013).
doi:10.1038/nature12190.


\bibitem{Fredrickson1984}
G. H. Fredrickson and H. C. Andersen,
``Kinetic Ising model of the glass transition,''
Phys. Rev. Lett. \textbf{53}, 1244--1247 (1984).
doi:10.1103/PhysRevLett.53.1244.

\bibitem{BagchiElastic2026}
B. Bagchi,
``Cooperative elastic mechanism of activated
structural relaxation in glassy liquids,''
J. Chem. Phys. \textbf{164}, 244503 (2026).
doi:10.1063/5.0330297.

\end{thebibliography}
\end{document}